\documentclass[12pt,preprint]{aastex}

\newcommand{\lsim}{\raise0.3ex\hbox{$<$}\kern-0.75em{\lower0.65ex\hbox{$\sim$}}}
\newcommand{\gsim}{\raise0.3ex\hbox{$>$}\kern-0.75em{\lower0.65ex\hbox{$\sim$}}}
\newcommand{\propsim}{\raise0.3ex\hbox{$\propto$}\kern-0.75em{\lower0.65ex\hbox{$\sim$}}}

\newcommand{\cmc}{\rm ~cm^{-3}}

\newcommand{\kms}{\rm ~km~s^{-1}}

\newcommand{\ml}{~M_\odot ~\rm yr^{-1}}

\def\EE#1{\times 10^{#1}}

\begin{document}
    
\title{Heating from free-free absorption and the mass-loss rate of the progenitor stars to supernovae}

\author{C.-I. Bj\"ornsson\altaffilmark{1} and P. Lundqvist\altaffilmark{1,2}}
\altaffiltext{1}{Department of Astronomy, AlbaNova University Center, Stockholm University, SE--106~91 Stockholm, Sweden.}
\altaffiltext{2}{The Oskar Klein Centre, AlbaNova, SE-106 91 Stockholm, Sweden}
\email{bjornsson@astro.su.se}
\email{peter@astro.su.se}

\begin{abstract}
An accurate determination of the mass-loss rate of the progenitor stars to core-collapse supernovae is often limited by uncertainties pertaining to various model assumptions. It is shown that under conditions when the temperature of the circumstellar medium is set by heating due to free-free absorption, observations of the accompanying free-free optical depth allow a direct determination of the mass-loss rate from observed quantities in a rather model independent way. The temperature is determined self-consistently, which results in a characteristic time dependence of the free-free optical depth. This can be used to distinguish free-free heating from other heating mechanisms. Since the importance of free-free heating is quite model dependent, this also makes possible several consistency checks of the deduced mass-loss rate. It is argued that the free-free absorption observed in SN 1993J is consistent with heating from free-free absorption. The deduced mass-loss rate of the progenitor star is, approximately, $10^{-5} \ml$ for a wind velocity of $10$\,km\,s$^{-1}$. 
\end{abstract}


\section{Introduction}
The properties of the circumstellar medium into which the explosion of a core-collapse supernova expands are determined by several different effects. The density structure is mainly set by the wind of the progenitor star, while its temperature is likely to be affected by radiation coming from the supernova. The initial flash of radiation from the breakout of the shock from below the stellar surface as well as the radiation from the viscous shocks resulting from the interaction with the circumstellar medium can both contribute to the heating. Since the resulting temperature depends also on the cooling properties of the gas, a self-consistent determination of the level of ionization is important. This is a non-trivial problem, which tends to make conclusions rather model dependent.

There are several ways to infer the density and temperature of the circumstellar medium. An analysis of the emission assumed to be emitted from behind the viscous shocks in the standard model \citep{che82a} can constrain its density. The external medium can also be studied directly either by the effects it has on the supernova radiation or its narrow line emission excited by this radiation \citep[e.g., SNe IIn;][]{fil97}.

The X-ray emission from the shocked gas is likely dominated by bremsstrahlung and/or Comptonization of the photospheric photons. A proper modeling of this emission should then give a good estimate of the density; i.e., the mass-loss rate of the progenitor star. In common with the temperature of the circumstellar medium, the result can be rather model dependent; for example, in SN 1993J both the forward \citep{fra96} and the reverse \citep{suz93} shock have been suggested as the location for the X-ray emission leading to very different mass-loss rates.

External free-free absorption is sensitive to the density and temperature of the circumstellar medium. It was recognized early on \citep{che82b} that the turn-over of the radio spectrum at low frequencies seen in several supernovae could, at least in part, be due to free-free absorption. With good enough data in the low frequency range, free-free absorption can be separated from  a possible contribution from synchrotron self-absorption. Even though the circumstellar medium can normally be assumed to be fully ionized \citep[e.g.,][]{che82b}, a limiting factor for those cases where free-free absorption is apparent is the sensitivity of the deduced mass-loss rate to its temperature. Usually this is dealt with by either assuming the temperature to have a given prescribed value \citep[e.g.,][]{wei02} or by parameterizing the temperature and deduce the relevant parameters from modeling \citep{f/b98}. In either case the accuracy of the mass-loss rate deduced from observations is no better than the assumptions made regarding the temperature.

This paper considers the possibility that the main heating mechanism of the circumstellar medium is free-free absorption. Observations of the resulting free-free optical depth then allows a self-consistent determination of the temperature and, hence, the density. It is argued that for supernovae expanding into a hydrogen rich circumstellar medium, free-free absorption may indeed set the temperature immediately ahead of the forward shock. In Section\,2 the heating due to free-free absorption of a synchrotron spectrum is discussed analytically using a few simplifying assumptions. The derived temperature is used in Section\,3 to calculate the resulting optical depth to free-free absorption. More exact numerical calculations are presented in Section\,4. The circumstances under which the initial temperature of the circumstellar medium is low enough for the heating by free-feee absorption to be important are discussed in Section\,5. A discussion of the results follows in Section\,6 together with a summery of the main conclusions.  Unless stated otherwise, cgs-units are used throughout the paper.

\section{Heating due to free-free absorption of a synchrotron spectrum} \label{sect2}

Free-free absorption leads to heating of the absorbing gas. Consider a fully ionized gas with density $n$, temperature $T$, and a radiation field described by an intensity $I(\nu,\theta)$. The heating can then be written
\begin{equation}
	\frac{{\rm d}E}{{\rm d}t} = \int {\rm d}\nu{\rm d}\Omega\,I(\nu,\theta)\,\mu(\nu),
	\label{eq:1.1}
\end{equation}
where $E=3nkT/2$ is the energy density of the gas and $\mu$ is the absorption coefficient. Most of the heating of the circumstellar gas occurs within one scale height in front of the forward shock, i.e., during a time $t \approx R/v_{\rm sh}$, where $R$ is the shock radius and $v_{\rm sh}$ is the shock velocity. The energy is absorbed by the electrons so that $T$ is the electron temperature. The effects of the nuclei are accounted for by introducing a factor $f_{\rm e}$ so that $E=3n_{\rm e}kTf_{\rm e}/2$, where $n_{\rm e}$ is the density of electrons. Here, $1\le f_{\rm e} \le 2$ and is determined by a combination of the transfer of energy from the electrons to the nuclei and the mass of the nuclei; e.g., $f_{\rm e} = 2$ for hydrogen and equipartition between electrons and protons, while $f_{\rm e} \approx 1$ for heavy nuclei and/or little transfer of energy to the nuclei.

Close to the shock, integrating over angles (${\rm d}\Omega$) gives, roughly, $2\pi$, while integration over frequency (${\rm d}\nu$) gives, roughly, $\nu$. Hence,
\begin{equation}
	\frac{3f_{\rm e}n_{\rm e}k}{2}\frac{{\rm d}T}{{\rm d}t} \approx 2\pi \nu I(\nu) \mu.
	\label{eq:1.2}
\end{equation}
The synchrotron spectrum is characterized by a self-absorption frequency $\nu_{\rm abs}$. Assume that the optical depth to free-free absorption at $\nu_{\rm abs}$ is less than unity, i.e., $\tau_{\rm ff}(\nu_{\rm abs}) < 1$. In this case most of the absorbed energy comes from around $\nu_{\rm abs}$. With $I(\nu_{\rm abs}) \approx 2kT_{\rm b}(\nu_{\rm abs}/c)^2$, where $T_{\rm b}$ is the brightness temperature, equation (\ref{eq:1.2}) leads to
\begin{equation}
	\frac{{\rm d}T}{{\rm d}t} \approx \frac{8\pi}{3}\frac{\nu_{\rm abs}^3 T_{\rm b} \mu}{c^2 n_{\rm e}
	f_{\rm e}}.
	\label{eq:1.3}
\end{equation}

In the absence of cooling, equation (\ref{eq:1.3}) can be integrated for a given mass element (i.e., $n_{\rm e} = $ constant). With $\mu = 1.8\EE{-2}\,g_{\rm ff}\,Z^2\,n_{\rm i}\,n_{\rm e}\,/T^{3/2}\,\nu^2$, the temperature of the gas at the shock front varies with time as
\begin{equation}
	T_{\rm sh} \approx 2.7\EE{5}\left(\frac{ g_{\rm ff}Z^2}{f_{\rm e}} n_{\rm i,8}\,\nu_{\rm abs,10}\,
	t_{\rm 10}\,T_{\rm b,11}\,\eta\right)^{2/5}.
	\label{eq:1.4}
\end{equation}
Here, $g_{\rm ff}$, $Z$, and $n_{\rm i}$ are, respectively, the free-free Gaunt factor, the charge and density of the ions. Furthermore $n_{\rm i,8} \equiv n_{\rm i}/(10^8 \cmc)$, $\nu_{\rm abs,10} \equiv \nu_{\rm abs}/(10^{10}\,{\rm Hz})$, $t_{\rm 10} \equiv t/(10\,{\rm days})$, and $T_{\rm b,11} \equiv T_{\rm b}/(10^{11} {\rm K})$. The factor $\eta$ accounts for the transition to the regime where heating is not important and is given by
 \begin{equation}
 	\eta = 1 + \left\{\frac{T_{\rm o}}{2.7\EE{5} \,\left(g_{\rm ff}Z^2 f_{\rm e}^{-1} n_{\rm i,8}\, 
	\nu_{\rm abs,10}\, t_{\rm 10}\,T_{\rm b,11}\right )^{2/5}}\right \}^{5/2},
	\label{eq:1.4a}
\end{equation}
 where $T_{\rm o}$ is the initial, constant temperature of the circumstellar medium. The radial variation above the shock front of the gas temperature at a given time can be calculated in a similar manner. 
 
 When the mass-loss rate of the progenitor star is constant, the electron density in the circumstellar medium is given by
 \begin{equation}
 	n_{\rm e} = 4.0\EE7\frac{\dot M_{\rm -5}}{v_{\rm w,1}\,v_{\rm sh,4}^2\,t_{\rm 10}^2\, 
	\mu_{\rm e}},
	\label{eq:1.5}
\end{equation}
where $\dot M$ and $v_{\rm w}$ are, respectively, the mass-loss rate and wind velocity of the progenitor star, $\mu_{\rm e}$ is the mean molecular weight of the electrons, and $R=v_{\rm sh}\,t$ has been used. Furthermore, $\dot M_{\rm -5} \equiv 1\EE{-5} \ml, v_{\rm w,1} \equiv  v_{\rm w}/(10\kms)$, and $v_{\rm sh,4} \equiv v_{\rm sh}/(10^4 \kms)$. This can be used in equation (\ref{eq:1.4}) to obtain
\begin{equation}
	T_{\rm sh} \approx 1.8\EE{5}\left(\frac{ g_{\rm ff}Z}{f_{\rm e}\,\mu_{\rm e}} \frac{\dot M_{\rm -5}\,
	\nu_{\rm abs,10}\,T_{\rm b,11}\,\eta}{v_{\rm w,1}\,v_{\rm sh,4}^2\,t_{\rm 10}}\right)^{2/5}.
	\label{eq:1.6}
\end{equation}

\section{The free-free optical depth} \label{sect3}

In situations when the main heating mechanism is free-free absorption, the resulting free-free optical depth can be determined self-consistently. Since both the density and the temperature of the electrons decrease away from the shock front, $\tau_{\rm ff} \approx \mu (R)\,R$. At the synchrotron self-absorption frequency, this yields
\begin{eqnarray}
	\tau_{\rm ff}(\nu_{\rm abs}) & \approx & 1.8\EE{-2}\frac{g_{\rm ff}Z\,n_{\rm e}^2(R)\,R}
	{\nu_{\rm abs}^2\,T_{\rm sh}^{3/2}} \nonumber\\
	  & \approx & 3.2 \frac{\left(Z\,g_{\rm ff}\right)^{2/5}\,f_{\rm e}^{3/5}}{\mu_{\rm e}^{7/5}\,
	  T_{\rm b,11}^{3/5}\,v_{\rm sh,4}^{9/5}\,t_{\rm 10}^{12/5}\,
	  \nu_{\rm abs,10}^{13/5}\,\eta^{3/5}}\left(\frac{\dot M_{\rm -5}}
	  {v_{\rm w,1}}\right)^{7/5},
	\label{eq:1.7}
\end{eqnarray}
where equations (\ref{eq:1.5}) and (\ref{eq:1.6}) have been used. For a constant brightness temperature, $\nu_{\rm abs} \propto B$, where $B$ is the magnetic field. The strength of the magnetic field is usually taken to scale inversely with either $t$ or $R$. Under such conditions equation (\ref{eq:1.7}) shows that $\tau_{\rm ff}(\nu_{\rm abs})$ varies only slowly with time; in particular, it may even increase.

With the use of equation (\ref{eq:1.7}), the temperature in equation (\ref{eq:1.6}) can also be expressed as
\begin{equation}
	T_{\rm sh} \approx 1.3\EE{5}\left(\frac{ g_{\rm ff}Z}{f_{\rm e}^2} \frac{\tau_{\rm ff}(\nu_{\rm abs})
	\,t_{\rm 10}\,\nu_{\rm abs,10}^4\,T_{\rm b,11}^2\,\eta^2}{v_{\rm sh,4}}\right)^{2/7}.
	\label{eq:1.7a}
\end{equation}
When an appreciable amount of free-free absorption (i.e., $\tau_{\rm ff}(\nu_{\rm abs})\,\lsim 1$) is present, equation (\ref{eq:1.7a}) shows the expected temperature of the gas to be around $10^5$\,K for typical supernova parameters.

It is seen from equation (\ref{eq:1.7}) that for a supernova shock expanding into a slow progenitor wind (i.e., $v_{\rm w,1} \sim 1$), $\tau_{\rm ff}(\nu_{\rm abs}) > 1$ is possible. When this is the case, the above discussion needs to be amended. For $\tau_{\rm ff}(\nu_{\rm abs}) > 1$, the heating rate is proportional to the absorption coefficient $\mu$, while the duration of the heating is roughly proportional to $\mu^{-1}$; hence, the energy absorbed is that for $\tau_{\rm ff}(\nu_{\rm abs}) \approx 1$. This gives a maximum value of the temperature, which is obtained from equation (\ref{eq:1.3}) and using $t \approx 1/\mu v_{\rm sh}$ instead of $t \approx R/v_{\rm sh}$.
\begin{eqnarray}
	T_{\rm sh,max} &\approx & \frac{20\pi\,\delta}{3\,f_{\rm e}}\frac{T_{\rm b}\,\nu_{\rm abs}^3}{
	n_{\rm e}\,v_{\rm sh}\,c^2} \nonumber\\
	& \approx & 5.8\EE4\frac{\mu_{\rm e}\, \delta}{f_{\rm e}}\frac{T_{\rm b,11}\,v_{\rm w,
	1}\,v_{\rm sh,4}\,t_{\rm 10}^2\,\nu_{\rm abs,10}^3}{\dot M_{\rm -5}} 
	\label{eq:1.8}
\end{eqnarray}
The factor $\delta$ takes into account the extra heating that occurs for an optically thin synchrotron spectrum with spectral index $\alpha < 1$, where $I(\nu) \propto \nu^{-\alpha}$. In this case, $\nu\,I(\nu)$ increases with frequency so that most energy is absorbed at a frequency $\nu_{\rm ff}$ such that $\tau_{\rm ff}(\nu_{\rm ff}) \approx 1$. Hence,
\begin{equation}
\delta \approx \left(\frac{\nu_{\rm ff}}{\nu_{\rm abs}}\right)^{1-\alpha} \approx \tau_{\rm ff}
	(\nu_{\rm abs})^{(1-\alpha)/2},	
	\label{eq:1.9}
\end{equation}
where $\tau_{\rm ff} (\nu)\,\propsim\,\nu^{-2}$ have been used.

\section{Numerical results} \label{sect4}
\begin{figure}
\plotone{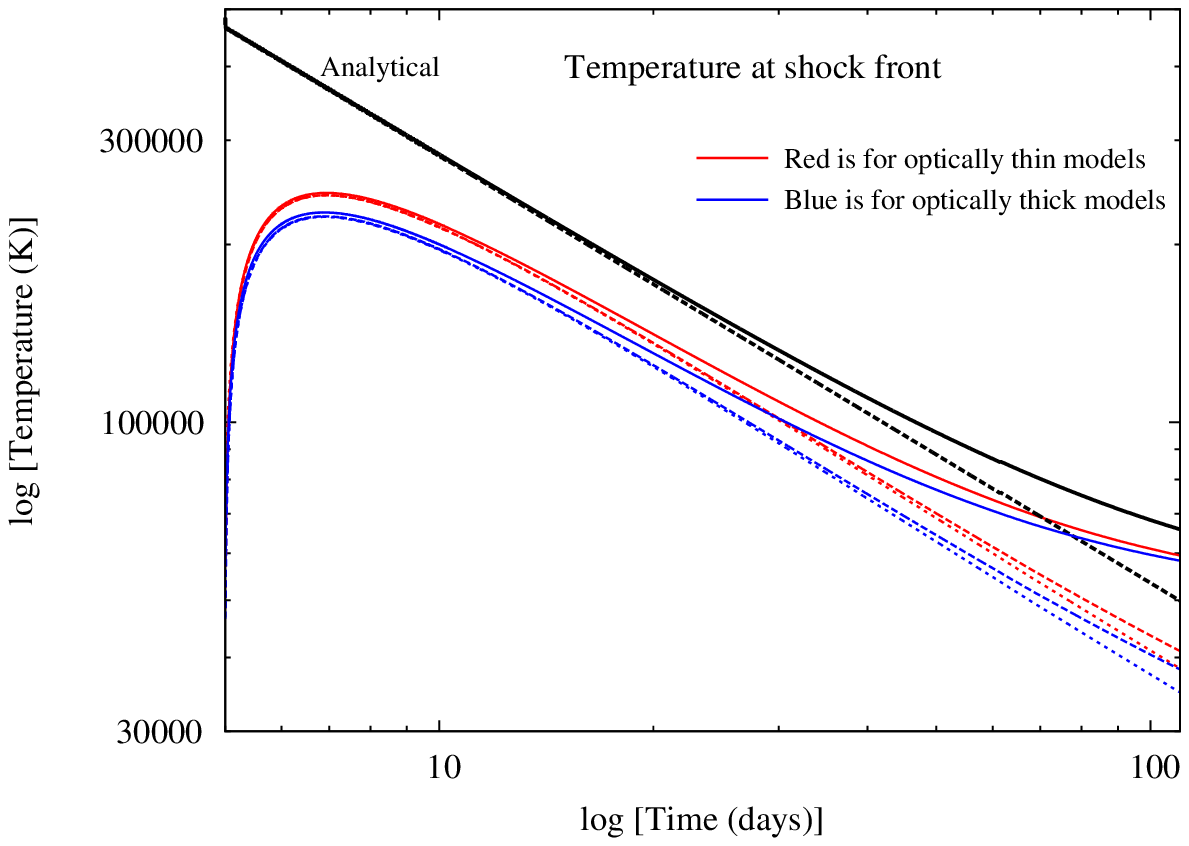}
\caption{The temperature of the circumstellar medium at the shock front is shown for $\dot M_{\rm -5}/v_{\rm w,1}=1$ and various initial temperatures as colored curves: 50\,000\,K (solid lines), 20\,000\,K (long-dashed lines), and 10\,000\,K (short-dashed lines). The curves labelled as "thin" correspond to the artificial case when the effects of the free-free absorption on the incident spectrum (see text) is neglected in the heating, while for the "thick" curves it is included (i.e., the physically relevant case). The black solid line is the analytical approximation in equation (\ref{eq:1.6})  for $T_{\rm o} = 50\,000$\,K and the black dashed line is for $T_{\rm o} = 0$\,K.
\label{fig1}} 
\end{figure}

\begin{figure}
\plotone{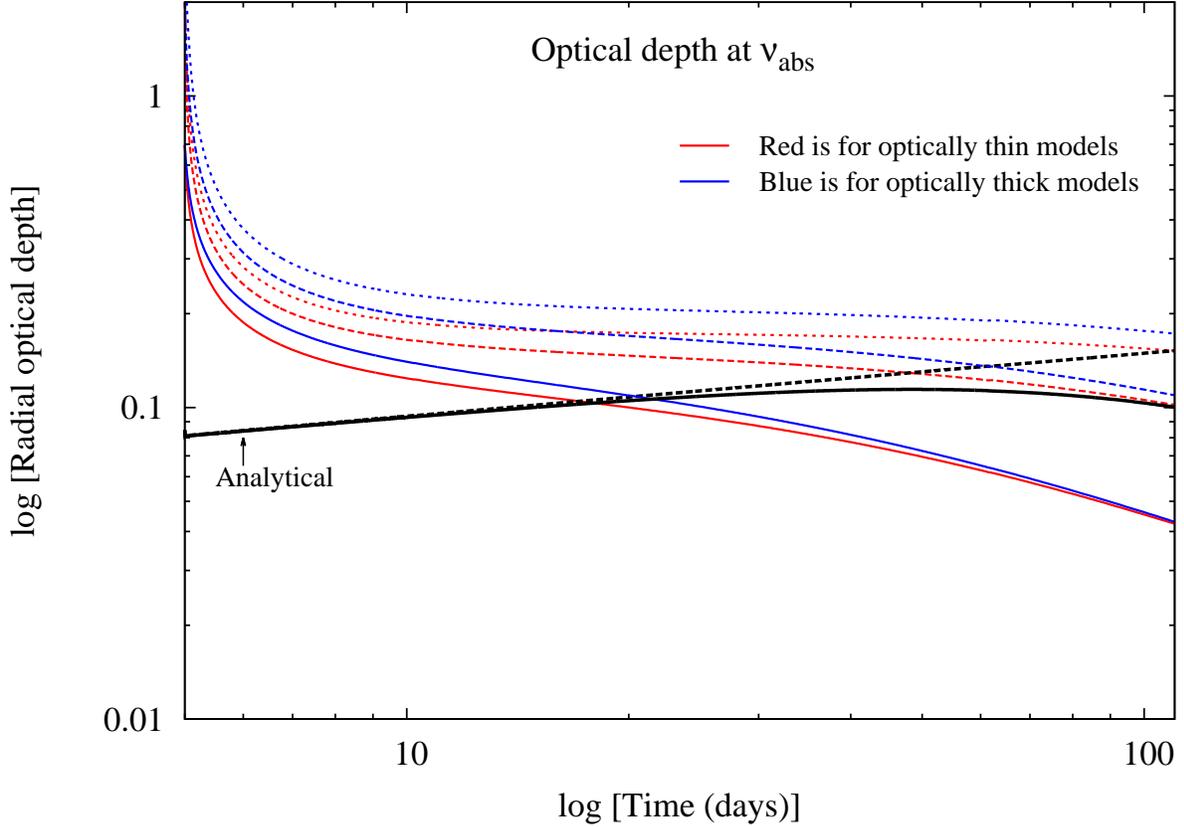}
\caption{The radial free-free optical depth of the circumstellar medium is shown for  $\dot M_{\rm -5}/v_{\rm w,1}=1$ and various initial temperatures of the circumstellar medium as colored curves: 50\,000\,K (solid lines), 20\,000\,K (long-dashed lines), and 10\,000\,K (short-dashed lines). The curves labelled as "thin" correspond to the artificial case when the effects of the free-free absorption on the incident spectrum (see text) is neglected in the heating, while for the "thick" curves it is included (i.e., the physically relevant case). The black solid line is the analytical approximation in equation (\ref{eq:1.7}) for $T_{\rm o} = 50\,000$\,K and the black dashed line is for $T_{\rm o} = 0$\,K.
\label{fig2}} 
\end{figure}

The synchrotron spectrum used in the numerical calculations is approximated by that coming from a spherical shell of radius $R$. Furthermore, the value of $\nu_{\rm abs}$ is assumed to be independent of the angle $\theta$. The intensity is then given by $I(\nu) = S(\nu) [1-\exp\{-{\tau_{\rm synch}(\nu)\}}]$, where $S(\nu) \propto \nu^{5/2}$ is the source function. Since $\mu \propto \nu^{-2}$, the main heating in equation (\ref{eq:1.2}) occurs somewhat below the frequency where the intensity peaks. In order to simplify the notation, $\nu_{\rm abs}$ is taken to be the frequency corresponding to $\tau_{\rm synch}(\nu_{\rm abs}) = 1$. The synchrotron optical depth is then given by $\tau_{\rm synch}(\nu) = (\nu/ \nu_{\rm abs})^{-(p+4)/2}$, where $N(\gamma) \propto \gamma^{-p}$ is the distribution of electron Lorentz factors $\gamma$. The peak intensity occurs at a synchrotron optical depth obtained from
\begin{equation}
	\exp \tau_{\rm synch}^{\rm peak} -1 = \frac{p+4}{5}\tau_{\rm synch}^{\rm peak},
	\label{eq:1.10}
\end{equation}
which corresponds to a frequency $\nu_{\rm peak}$. The brightness temperature is defined by the intensity at $\nu_{\rm peak}$; i.e., $I(\nu_{\rm peak}) \equiv 2kT_{\rm b}(\nu_{\rm peak}/c)^2$. The synchrotron spectrum can then be written as
\begin{equation}
	I = \frac{2kT_{\rm b}}{f(\nu_{\rm peak}/{\nu_{\rm abs}})c^2} \frac{\nu^{5/2}}{\nu_{\rm abs}^{1/2}}
	[1-\exp \{-\left(\frac{\nu_{\rm abs}}{\nu} \right)^{(p+4)/2}\}],
	\label{eq:1.11}
\end{equation}
where $f(x) = x^{1/2}[1-\exp \{-x^{-(p+4)/2}\}]$. In the calculations below $p=2$, which implies $x = 1.413$ and, hence, $f=0.355$. It may be noted that the brightness temperature at $\nu_{\rm abs}$ is larger than that at $\nu_{\rm peak}$ by a factor $(1-e^{-1})/f = 1.78$. In order to illustrate the accuracy of the analytical approximations as compared to the numerical results, in Figures (\ref{fig1}) and (\ref{fig2}) this correction factor has been applied to the brightness temperature of the former.

With the use of this intensity, the heating of the circumstellar gas is integrated in time as given by equation (\ref{eq:1.1}). In the numerical models, we take into account the angle-averaged optical depth for the attenuation of the intensity, and not just the radial optical depth as in equation (\ref{eq:1.7}). The gas consists of hydrogen and helium, which are assumed to be fully ionized. In addition to the heating, we also include free-free and Case B free-bound cooling as given by \cite{fer92}. The initial temperature of the circumstellar gas, when the free-free heating sets in, is a free parameter. The mass-loss rate of the progenitor star is assumed to be constant, which, in the standard self-similar model \citep{che82a}, implies that the radius of the forward shock varies as $R\propto t^{(n-3)/(n-2)}$, where $n$ is the power law index of the density structure of the supernova ejecta. 

In Figures (\ref{fig1}) and (\ref{fig2}) the numerical results for $n=30$ are compared to the analytical approximations in Sections 2 and 3. The heating is assumed to start instantaneously at $5$\,days and the scaling of $R$ is such that it coincides with the value used in \cite{f/b98} at $10$\,days. Furthermore, $T_{\rm b,11} = 1$, $v_{\rm w,1}=1$ and, for simplicity, the gas is assumed to consist of hydrogen only.  Since the equipartition timescale \citep[e.g.,][]{spi78} for the temperatures and densities implied by equations (\ref{eq:1.4}) and (\ref{eq:1.5}) is much shorter than the dynamical timescale, $f_{\rm e} =2$ is used. For the chosen parameters cooling has a minor effect on the temperature and was not included. More detailed calculations are presented in Section\,\ref{sect5b1} in which both cooling and helium are included. The radial optical depth shown in Figure\,(\ref{fig2}) corresponds to the center of the source. The synchrotron self-absorption frequency is an important parameter. Its value is taken to coincide with that derived in \cite{f/b98} at $10$\,days and vary with radius as $\nu_{\rm abs} \propto 1/R$. Anticipating the discussion in Section\,\ref{sect5b1} of SN 1993J, the parameter values have been chosen close to those deduced in \cite{f/b98}. For other parameter values the results can be obtained from Figures\,(\ref{fig1}) and (\ref{fig2}) by using equations (\ref{eq:1.6}) and (\ref{eq:1.7}) as scaling relations.

It is seen in Figure\,(\ref{fig1}) that the main heating occurs over one scale height ahead of the shock so that it takes, roughly, one dynamical time after the heating has started before the temperature immediately in front of the shock has reached the regime where equation (\ref{eq:1.6}) is valid. Except for the initial rise in temperature, which was not included in Sections 2 and 3, the agreement between the analytical and numerical results for the temperature and free-free optical depth is rather good. Although the analytical approximations  give a somewhat too high a temperature, its variation with time is well described by the analytical expressions. It may be noticed that neglecting the correction factor for the analytical results discussed above give good agreement also for the amplitude of the temperature. The analytical approximations use the temperature at the shock to calculate the free-free optical depth. This overestimates the average temperature above the shock and, hence, gives a value of the free-free absorption which is too low. As the temperature at the shock approaches $T_{\rm o}$ the relative importance of the radial variation of the density increases. Since the analytical result (equ.\,(\ref{eq:1.7})) assumes constant density within one scale height above the shock, it overestimates the average density, which then leads to an overestimate of the free-free optical depth at later times. For the case shown in Figure (\ref{fig2}), these two effects together cause the decline of the free-free optical depth with time to be underestimated. 

Another effect not included in Sections 2 and 3 is the attenuation of the incident intensity by the free-free absorption itself; i.e., the circumstellar medium is artificially assumed to be optically thin. This overestimates the heating and, hence, the temperature. In order to illustrate the effects of the free-free optical depth, the heating has also been calculated by artificially assuming the incident spectrum to be unaffected by free-free absorption. The result is shown as the optically thin models in Figures (\ref{fig1}) and (\ref{fig2}). Figure\,(\ref{fig2}) shows that although the optical depth in this case is lower, which is due to the higher temperature, its variation with time is similar to the physically realistic situation. The reason is that, in this case, the value of $\tau_{\rm ff}(\nu_{\rm abs})$ varies only slowly with time so that the scaling between the optically "thin" and "thick" cases is roughly constant. In general, when the optical depth is large enough to affect the heating, variations of, for example, $\nu_{\rm abs}$ and $v_{\rm sh}$ will influence the time evolution of $\tau_{\rm ff}(\nu_{\rm abs})$ (cf. equ.\,(\ref{eq:1.7})). When $\tau_{\rm ff}(\nu_{\rm abs})$ decreases (increases) with time, the average heating ahead of the shock decreases slower (faster) than for the optically "thin" case. This leads to a slower (faster) decline of the average temperature and, hence, to a steeper (flatter) decline of the free-free absorption.

\section{The initial temperature of the circumstellar medium}\label{sect5}
As discussed in Section\,3, for typical supernova parameters the maximum temperature of the circumstellar medium due to heating from free-free absorption is around $10^5$\,K and occurs for optical depths of order unity (cf. equ.\,(\ref{eq:1.7a})). Hence, in order for free-free absorption to be an important heating mechanism, the heating of the circumstellar medium prior to the radio emitting phase must not give temperatures higher than this.

\subsection{Heating by the radiation from the breakout of the radiation mediated shock}\label{sect5a}
The breakout of the radiation mediated supernova shock from below the stellar surface is expected to give rise to an initial flash of energetic radiation that is able to ionize and heat the surrounding medium. Except for low shock velocities at breakout, $ v_{\rm s}\,\lsim\,0.02\,c$ \citep{wea76,sap13}, the radiation is likely to deviate from black body. The calculations by \cite{sap13} assumed the photons to be produced by bremsstrahlung and that local Compton equilibrium was reached by all photons which were not absorbed. However, close to the breakout, there is also the possibility for the bremsstrahlung photons to escape. Although energetically these escaping photons are likely to be unimportant, their low frequencies could make them dominate the ionization and, hence, the temperature of the circumstellar medium at distances large enough for recombination not to occur before reached by the viscous shock. This requires that they are not all absorbed as they diffuse out from behind the shock at breakout.

Let $h\nu_{\rm I} \ll kT_{\rm Comp}$, where $\nu_{\rm I}$ is the ionization frequency of hydrogen and $T_{\rm Comp}$ is the temperature behind the shock. The number of bremsstrahlung photons with frequencies $\approx \nu_{\rm I}$ produced per unit area per second within a distance $\Delta r$ behind the shock is
\begin{equation}
	N(\nu_{\rm I}) \approx \left(\frac{8}{3\pi}\right)^{1/2}\left(\frac{mc^2}{kT_{\rm Comp}}\right)^{1/2}
	c\,\alpha_{\rm f}\, \sigma_{\rm T}\, n_{\rm s,e}\, n_{\rm s,i}\, g_{\rm ff} Z^2 \Delta r,
	\label{eq:1.12}
\end{equation}
where $\alpha_{\rm f}$ is the fine structure constant, $\sigma_{\rm T}$ is the Thomson cross section,  $n_{\rm s,e}$ and $n_{\rm s,i}$ are, respectively, the densities of electrons and ions behind the shock. In a steady state situation, the minimum distance ($\Delta r_{\rm I}$) ahead of the shock that can be ionized is then obtained by assuming negligible contribution to the ionization from other sources, e.g., collisions,
\begin{equation}
	N(\nu_{\rm I}) \approx n_{\rm e}\, n_{\rm p}\, \beta\, \Delta r_{\rm I},
	\label{eq:1.13}
\end{equation} 
where $n_{\rm p}$ is the density of protons and $\beta$ is the recombination coefficient. By introducing the corresponding scattering optical depths, $\tau_{\rm b} \equiv \sigma_{\rm T}\, n_{\rm s,e}\, \Delta r$ and $\tau_{\rm I} \equiv \sigma_{\rm T}\, n_{\rm e}\, \Delta r_{\rm I}$, equation (\ref{eq:1.13}) can be written
\begin{equation}
	\frac{\tau_{\rm I}}{\tau_{\rm b}} \approx 1.0\,\frac{g_{\rm ff}}{T_{\rm Comp,6}^{1/2}
	\beta_{\rm -14}}\, \frac {Z^2\, n_{\rm s,i}}{n_{\rm p}},
	\label{eq:1.14}
\end{equation}
where $T_{\rm Comp,6} \equiv T_{\rm Comp}/(10^6\, {\rm K})$ and $\beta_{\rm -14} \equiv \beta / (10^{-14}{\rm cm^3\,sec^{-1}})$. In a radiation mediated shock and a medium dominated by hydrogen and helium $Z^2\,n_{\rm s,i}/n_{\rm p} = 7(1+Y)$, where $Y$ is the mass fraction of helium. Since the number of escaping photons is much higher than the number of electrons ahead of the shock, these photons are expected to heat the electrons to a temperature similar to $T_{\rm Comp}$ without changing the emerging spectrum significantly \citep{sap13}. As shown by \cite{sap13}, the value of $T_{\rm Comp}$ is in the range $10^6 - 10^7$\,K for $v_{\rm s}/c \approx 0.1$. With $\beta_{\rm -14} \approx 1$ for $T_{\rm Comp,6} \approx 1$ and $g_{\rm ff} \sim 1$, it is seen from equation (\ref{eq:1.14}) that $\tau_{\rm I}\, \gsim\, 7 \tau_{\rm b}$. The weak temperature dependence of $T_{\rm Comp}^{1/2} \beta$ implies that $\tau_{\rm I} > \tau_{\rm b}$ should apply to most shock breakouts independent of supernova type.

The value of $\tau_{\rm b}$ is determined by the condition that the Compton parameter is roughly unity, i.e.,
\begin{equation}
	\frac{4kT_{\rm Comp}}{mc^2} \tau_{\rm b}^2 \approx 1.
	\label{eq:1.15}
\end{equation}
The temperature immediately behind the shock is determined by the density of photons. In a steady state situation this is roughly determined by the photons produced within one diffusion length. The resulting value of $T_{\rm Comp}$ increases very rapidly with $v_{\rm s}$. For an accelerating shock close to breakout, this corresponds to the number of photons produced during one dynamical time. The velocity gradient behind the shock gives in this case also a contribution of photons diffusing up to the shock front from regions downstream with higher photon density. The relative importance of this latter contribution depends on the deviation from LTE behind the shock, which, in turn, increases with shock velocity. As shown by \cite{sap13}, photon diffusion causes the variation of  $T_{\rm Comp}$ with $v_{\rm s}$ to be more moderate. Since the scattering optical depth at shock breakout is $\tau_{\rm s} \approx c/v_{\rm s}$, it is interesting to note that $T_{\rm Comp}/v_{\rm s}^2$ increases only slowly with $v_{\rm s}$ \citep[see Fig.\,3 in][]{sap13}; hence, $\tau_{\rm b}/\tau_{\rm s}$ decreases slowly with $v_{\rm s}$. Furthermore, $\tau_{\rm b}\, \gsim \,\tau_{\rm s}$ for typical supernova parameters. The value of $T_{\rm Comp}$ decreases with decreasing density, since the number of available photons increases due to a lower free-free absorption frequency. Only for the highest densities, corresponding to Wolf-Rayet progenitor stars, is $\tau_{\rm b}\, \approx \,\tau_{\rm s}$.

The bremsstrahlung photons, which escape before reaching Compton equilibrium, are located right behind the shock front and, hence, are part of the initial phase of the shock breakout flash. The fraction that is absorbed while diffusing out is approximately $\tau_{\rm s}/\tau_{\rm I}$. From equation (\ref{eq:1.14}) and the discussion above, for progenitor stars with a hydrogen dominated atmosphere this is at most 10\%. The main point is that the initial phase of the flash has a significant fraction of photons with a bremsstrahlung spectrum. Hence, at a distance where the recombination time is longer than the dynamical time of the viscous shock, the temperature of the circumstellar medium is determined by ionization with a bremsstrahlung spectrum. For a power law distribution of photons with spectral index $\alpha$ and a photoionization cross-section varying as $\nu^{-3}$, the average photoionization frequency is $(\alpha+3)/(\alpha+2)\nu_{\rm I}$. The corresponding temperature for a bremsstrahlung spectrum ($\alpha = 0)$ is $7.9\times 10^4$\,K. A more accurate treatment of the frequency dependence of the photoionization cross-section gives a temperature somewhat lower; for example, \cite{lun92} finds 5\,-\,6\,$\times 10^4$\,K for a flat spectrum.

\subsection{Heating by radiation from the viscous shock}\label{sect5b}
The viscous shock propagating into the circumstellar medium produces high energy radiation. This will Compton scatter and thereby heat the circumstellar medium ahead of the shock if $\alpha \,\lsim \,2$. The most likely emission mechanisms of this radiation is bremsstrahlung and Comptonization of the photospheric photons. Both of these are sensitive to density and, hence, to $\dot M / v_{\rm w}$. The resulting temperature is rather model dependent, since its value depends on cooling, which, in turn, is determined by the ionization. When this heating is important the rise in temperature ahead of the shock is likely to be most prominent at small radii; hence, the temperature is expected to decline with radius and at some point there should be a transition to the constant temperature set by the radiation from the breakout shock. This assumes a fully ionized circumstellar medium, which may not be achieved for large mass-loss rates \citep[e.g.,][]{lf88}.

\subsubsection{Application to SN 1993J}\label{sect5b1}
The radio spectrum from SN 1993J showed clear evidence for free-free absorption during the first 100\,days \citep{f/b98,per01}. In \cite{f/b98} this was modeled by parameterizing the temperature variation as an initial decline with a later transition to a constant value. As compared to the temperature structure calculated in \cite{fra96}, the deduced parameters indicated a much slower initial decline and a transition to the constant temperature part at a considerable later date. The rapid temperature decline in \cite{fra96} results in a roughly constant free-free optical depth at a given frequency ($\tau_{\rm ff}(\nu)$) until day\,15 after which the constant temperature range caused it to decline approximately as $t^{-3}$. The free-free absorption derived in \cite{f/b98} for SN 1993J instead indicates a transition to a constant temperature at $\approx 80$\,days. However, the deduced parameters are such that an approximate power law dependence is obtained over the whole observed time interval with $\tau_{\rm ff}(\nu)\, \propsim\, t^{-2.1}$.

As mentioned above, the temperature structure is sensitive to model assumptions; hence, these discrepancies do not necessarily compromise the conclusions drawn in \cite{fra96} and \cite{f/b98}. However, since the temperature of the circumstellar medium heated by free-free absorption declines considerably slower (see Fig.\,\ref{fig1}) than that deduced in \cite{fra96}, it is interesting to compare the time variation of $\tau_{\rm ff}(\nu)$ expected in the two cases. Since $\tau_{\rm ff}(\nu) = \tau_{\rm ff}(\nu_{\rm abs}) (\nu_{\rm abs}/\nu)^2$ it is seen from equation (\ref{eq:1.7}) that  $\tau_{\rm ff}(\nu) \propto t^{-12/5}\, \nu_{\rm abs}^{-3/5}\, v_{\rm sh}^{-9/5}\,\eta^{-3/5}$. From \cite{f/b98} one finds $\nu_{\rm abs} \propto t^{-0.68}$ and $v_{\rm sh} \approx $\,constant up to day\,100, which leads to $\tau_{\rm ff}(\nu) \propto t^{-2.0}\,\eta^{-3/5}$. Hence, neglecting for the moment the effects of a finite initial temperature of the circumstellar medium, this is consistent with that found in \cite{f/b98}. The implied mass-loss rate is then significantly smaller than assumed in \cite{f/b98}; for example, using the same wind velocity gives a mass-loss rate roughly a factor of five lower (i.e., $\dot M_{\rm -5} \approx 1$). 

It can be seen from the calculations in \cite{fra96} that for such a low mass-loss rate, the heating of the circumstellar medium by the radiation produced behind the viscous shock may become negligible \citep[cf.][for models with $\dot M_{\rm -5} \lsim 1$]{lun13}. Furthermore, the recombination timescale for hydrogen becomes longer than the dynamical time of the viscous shock at around 10\,days, which corresponds to the beginning of radio observations. Hence, the initial temperature of the circumstellar medium at later times could be set by the emission from the breakout of the radiation mediated shock. The time for the transition in equation (\ref{eq:1.7}) to a circumstellar medium with constant temperature (i.e., $\tau_{\rm ff}(\nu) \propto t^{-3}$) depends on the initial temperature. Figure\,(\ref{fig1}) shows that a transition at $\approx 80$\,days, as deduced for SN 1993J, corresponds roughly to an initial temperature of 50\,000\,K. This is consistent with  ionization and heating by a bremsstrahlung spectrum from the initial flash of radiation from the breakout of the supernova shock (cf. Sec.\,5.1). 

\begin{figure}
\plotone{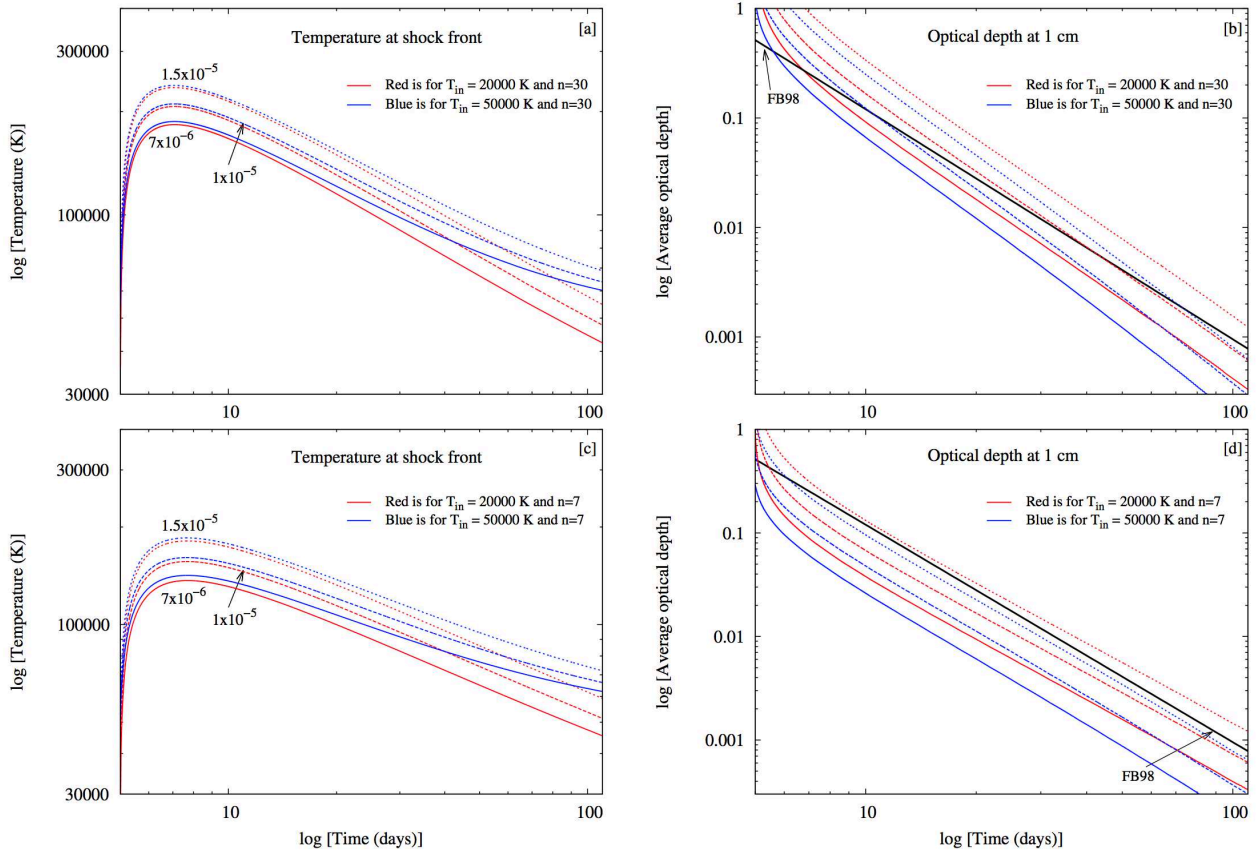}
\caption{The temperature of the circumstellar medium at the shock front and the average free-free optical depth are shown for various mass-loss rates in units of $\ml$ (assuming a wind velocity of $10$\,km\,s$^{-1}$) and initial temperatures of the circumstellar medium. The two different models (labelled $n=7$ and $n=30$, respectively) are scaled such that their radii are the same at $100$\,days. The black solid line in panels (b) and (d) corresponds to the approximate expression deduced in \cite{f/b98} for the free-free optical depth in SN 1993J. The mass-loss rates for the various optical depth curves in panels (b) and (d) are the same as those indicated for the corresponding temperatures in panels (a) and (c).
\label{fig3}} 
\end{figure}

Although cooling is only marginally important for $\dot M_{\rm -5} \sim 1$, it is seen from Figure\,(\ref{fig2}) that optical depth effects are significant, which leads to somewhat lower temperatures. A self-consistent inclusion of the free-free optical depth is therefore needed in order to derive a reliable estimate of the mass-loss rate. This is done in the calculations presented in Figure\,(\ref{fig3}), where, also, a helium-to-hydrogen ratio of 0.3 \citep{shi94} is used and, for completeness, cooling is included. The free-free optical depth in Figures\,(\ref{fig3}b) and (\ref{fig3}d) is now angle-averaged over the source.

It is seen in Figure\,(\ref{fig3}b) that the agreement between the average decline of the free-free optical depth in SN 1993J and the analytical approximations is fortuitous for $n=30$. The transition to the constant temperature regime of the circumstellar medium steepens the decline to the extent that it no longer is consistent with that deduced by \cite{f/b98}. If this average decline reflects the true properties of the circumstellar medium, it would imply a temperature gradient larger than expected for free-free heating alone. It could be caused by a small amount of additional heating corresponding to the beginning of the radio phase; for example, due to X-ray emission from the viscous shocks  and/or from the breakout shock.

However, the usefulness of a detailed comparison between the deduced average time variation of  $\tau_{\rm ff}(\nu)$ with that expected in a free-free heating scenario may be limited by the sensitivity to the shock velocity. Appreciable free-free absorption in SN 1993J was observed to be present mainly during the initial phase when the measured shock velocity was constant (roughly, the first 100\,days). The transition to a constant temperature of the circumstellar medium at $\approx 80$\,days, which led to a steeper time dependence of $\tau_{\rm ff}(\nu)$, was compensated by the start of a decreasing shock velocity at 100\,days. Hence, the value of the exponent in the approximate relation $\tau_{\rm ff}(\nu)\, \propsim\, t^{-2.1}$ in \cite{f/b98} is influenced by the shock velocity during the time when no free-free absorption could be measured. 

In the discussion above it has tacitly been assumed that the shock velocity also corresponds to the observed expansion velocity of the outer boundary of the synchrotron source. This is not necessarily the case. Relaxing this assumption gives an alternative way to reconcile the observed time variation of $\tau_{\rm ff}(\nu)$ with that expected from a free-free heating scenario. It is seen from equation (\ref{eq:1.7}) that a shock velocity decreasing with time results in a slower variation of the optical depth. Observations by \cite{bru10} and \cite{bie11} show that the velocity of the outer boundary of the synchrotron source varies with time roughly as $t^{-0.8}$ after $100$\,days, which corresponds to $n=7$ in the standard self-similar model. Figures\,(\ref{fig3}c) and (\ref{fig3}d) show the temperature at the shock front and the optical depth for a model with $n=7$ also during the first $100$\,days. In this case the scaling is such that the radius at $100$\,days is the same as the value used in \cite{f/b98}. It is seen that the variation of the optical depth with time agrees well with that expected for free-free heating; in particular, the agreement is better with an initial temperature of the circumstellar medium corresponding to $50\,000$\,K rather than $20\,000$\,K. It may be noticed that the slopes of the various curves in Figures\,(\ref{fig3}b) and (\ref{fig3}d) are independent of the mass-loss rate. Instead  they are determined by the combined time variations of the shock velocity, the synchrotron self-absorption frequency and the transition to the initial, constant temperature of the circumstellar medium ($T_{\rm o}$). Once the slope agrees with observations, the mass-loss rate is obtained from the amplitude of the free-free absorption (cf. equ.\,(\ref{eq:1.7})).

The value of the mass-loss rate deduced from Figure\,(\ref{fig3}d) depends on the brightness temperature. The brightness temperature observed in SN 1993J was considerably lower than expected for a standard homogeneous source. In the model by \cite{f/b98} this was accounted for by a combination of a magnetic field much stronger than the equipartition value and cooling of the relativistic electrons. Together they lowered the brightness temperature by roughly a factor of three, with the contribution from the latter slowly decreasing with time due to the decreasing importance of cooling. If the temperature of the circumstellar medium in SN1993J is set by free-free heating, its density is substantially smaller than assumed in \cite{f/b98}. As a result the low brightness temperature is unlikely due only to a strong magnetic field and/or cooling. A direct way of lowering the brightness temperature is by invoking inhomogeneities \citep[e.g.,][]{bjo13}. However, even for a homogeneous circumstellar medium, the heating could be patchy due to a varying local brightness temperature. The use of an average value of the brightness temperature in such situations may limit the accuracy to which the free-free optical depth can be modeled.

The most accurate estimate of the mass-loss rate is, therefore, obtained by comparing the free-free optical depths at $100$\,days, when the observed brightness temperature is only a factor $2\,-\,3$ smaller than the standard value. Due to scaling of the radius, the two models ($n=7$ and $n=30$) shown in Figure\,(\ref{fig3}) give approximately the same result at this time. With a standard brightness temperature somewhat lower than $10^{11}$\,K,  the mass-loss rate should be estimated using an effective brightness temperature $T_{\rm b,11} \approx 0.2\,-\,0.3$. For a given curve of the average optical depth in Figures\,(\ref{fig3}b) and (\ref{fig3}d), the mass-loss rate and brightness temperature are related by $\dot M_{\rm -5} \propto T_{\rm b,11}^{3/7}$ (cf. equ.\,(\ref{eq:1.7})). Since the best fit corresponds to  $\dot M_{\rm -5} \approx 1.6$ ($v_{\rm w,1}=1$) for $T_{\rm b,11} =1$, correcting for the value observed in SN 1993J then yields $\dot M_{\rm -5} \approx 0.8\,-\,1.0$.

Although the heating mechanism of the circumstellar medium in SN 1993J cannot be definitely established, it is possible to give a roughly consistent description of the temperature variation ahead of the shock as due to heating by free-free absorption of a circumstellar medium with an initial temperature of  $\sim$\,50\,000\,K. The implications of this for the deduced properties of the viscous shock will be discussed in a forthcoming paper.

\section{Discussion and conclusions}\label{sect6}
When the temperature of the circumstellar medium is determined by free-free absorption, observation of the free-free optical depth makes it possible to directly relate the mass-loss rate to observed quantities only. This gives a method to obtain a value of the mass-loss rate which is considerably less model dependent than those normally used. It should be noticed that the brightness temperature is often not directly observable. Although for a standard synchrotron model its value is quite insensitive even to rather large variations of source parameters, the presence of inhomogeneities, large deviations from equipartition between magnetic fields and relativistic particles and/or cooling of the latter can have non-negligible effects on the deduced mass-loss rate. 

The analytical expression for the mass-loss rate (equ.\,(\ref{eq:1.7})) gives a rather good estimate for moderate optical depths. Although the temperature immediately ahead of the shock is adequately described by the analytical expressions, increasing the optical depth causes the heating to be localized closer to the shock. This leads, on average, to a lower temperature in the region where most of the free-free absorption occurs. Hence, using the analytical expression for the mass-loses rate would then give a value larger than the actual one. 

Since the time variation of the free-free optical depth is expected to be the best way to distinguish free-free absorption from other heating mechanisms, these analytical approximations should allow to evaluate the importance of the former. They can also be used to make consistency checks of the deduced mass-loss rate. For large mass-loss rates several effects can invalidate one or more of the assumptions leading up to the results in Sections 2 and 3; for example, at some density cooling will balance free-free heating and thereby reduce the maximum temperature of the circumstellar gas ahead of the shock. Likewise, recombination may become important at distances corresponding to the initial phases of radio emission. When this occurs, reionization by radiation with a spectrum hard enough would result in a temperature larger than that possible for free-free heating, making free-free heating negligible. Also, the Comptonized radiation from behind the shock hardens with increasing density and could provide an additional heating mechanism \citep{fra96}. The mass-loss rates, where these various effects set in, are model dependent and, in particular, the shock velocity is important.

A prerequisite for free-free heating to leave a distinct mark on the free-free absorption is an initial temperature not much larger than $10^5$\,K. As discussed above, this requires an ionizing spectrum that is not too hard; for example, a black body or a Wien spectrum would give too high a temperature. Hence, observations of free-free absorption can be used to constrain the spectral distribution of the radiation in the initial flash associated with the breakout of the supernova shock. It was argued in Section 5.2.1 that the free-free absorption in SN 1993J could plausibly be explained by a phase where free-free heating was important. This would imply that a significant fraction of the bremsstrahlung seed photons behind the breakout shock did not reach Compton equilibrium.

The mass-loss rate of the progenitor stars to supernovae is an essential quantity not only for understanding the later evolutionary stages of massive stars but also as an input parameter for the physics governing the conditions behind the viscous shock. 

The non-thermal properties of the shocked gas are an important aspect particularly of the forward shock. These are not well known. The reason is not only limited observations but also that the relevant physics is only partly understood; for example, the injection problem \citep[e.g.,][]{bla94} is still not solved and the amplification of the magnetic field behind shocks cannot yet be calculated from basic physics. The determination from observations of the fractions of the thermal energy input behind the shock which go into relativistic electrons and magnetic fields could give important constraints to ongoing attempts to understand both of these processes \citep [e.g.,][]{c/s13,ell13}. Although, in principle, the energy densities of relativistic electrons and magnetic fields can be obtain from an analysis of the synchrotron emission, their fraction of the total energy density behind the shock depends on the density of the circumstellar medium, i.e., the mass-loss rate of the progenitor star. A related issue is the occurrence of non-linear shocks \citep[eg.,][]{b/e99,bla05} in which the pressure behind the shock is dominated by relativistic particles. Again, the efficiency of injecting particles into the acceleration process is an important parameter determining the characteristics of such shocks. A reliable determination of the mass-loss rate could then contribute to a better understanding of a wide range of issues. 

In conclusion, the main points of the present paper can be summarized as follows:

1) When the heating of the circumstellar medium is by free-free absorption, its temperature can be calculated self-consistently. As a result the density of the circumstellar medium can be obtained directly from the observed free-free optical depth in a rather model independent way. The maximum temperature is around $10^5$\,K for typical supernova parameters and occurs for optical depths of order unity.

2) It was argued that for a hydrogen rich circumstellar medium, the temperature resulting from ionization by the initial flash of radiation from the shock breakout is lower than $10^5$\,K. Hence, for such supernovae the main heating mechanism of the circumstellar medium may be free-free absorption. 

3) The observed time dependence of the free-free optical depth in SN 1993J can be accounted for by heating due to free-free absorption. The resulting value of the mass-loss rate of the progenitor star is estimated to be $(0.8\,-\,1.0)\times 10^{-5}\ml$ for a wind velocity of $10$\,km\,s$^{-1}$.

P.L. acknowledges support from the Swedish Research Council.

\clearpage

\end{document}